\documentclass[sigplan]{acmart}

\usepackage{url}
\usepackage{amsmath}
\usepackage{float}
\usepackage{enumitem}
\usepackage{color}
\usepackage{caption}
\usepackage{subcaption}
\usepackage{booktabs}
\usepackage{graphicx}
\usepackage{multicol}
\usepackage{tikz-cd}
\usepackage{svg}
\usepackage{xspace}

\NewEnviron{myequation}{%
    \begin{equation*}
    \scalebox{1.6}{$\BODY$}
    \end{equation*}
}

\newcommand{\systemName}{\texttt{CASPER}\xspace}

\renewcommand\footnotetextcopyrightpermission[1]{}
\setcopyright{acmcopyright=false}

\copyrightyear{2023}
\acmYear{2023}
\setcopyright{acmlicensed}
\acmConference[IGSC '23]{THE 14th international Green and Sustainable Computing Conference}{October 28--29, 2023}{Toronto, ON, Canada}
\acmBooktitle{THE 14th international Green and Sustainable Computing Conference (IGSC '23), October 28--29, 2023, Toronto, ON, Canada}

\title{CASPER: Carbon-Aware Scheduling and Provisioning for Distributed Web Services}
\titlenote{This work appeared at IGSC'23.}

\author{Abel Souza, Shruti Jasoria, Basundhara Chakrabarty, David Irwin, Prashant Shenoy}
\affiliation{%
  \institution{University of Massachusetts Amherst}
  \country{USA}
}

\author{Alexander Bridgwater, Axel Lundberg, Filip Skogh, Ahmed Ali-Eldin}
\affiliation{%
  \institution{Chalmers University of Technology}
  \country{Sweden}
}

\date{October 28--29, 2023}

\begin{document}
\begin{abstract}
%It is of utmost importance to actively reduce carbon costs and transition towards more sustainable energy sources, particularly in the field of computing. However, there is lack of visibility into carbon costs at the data center level, which impedes operators from efficiently optimizing their carbon efficiency.
%Computation often has significant spatial, temporal, and performance flexibility, enabling it to shift the location, time, and intensity of its execution to better align with the availability of carbon-free renewable energy or low-carbon grid energy. A prime example of this is a geographically distributed web application hosted across various cloud regions worldwide, each with distinct carbon costs depending on their electricity sources. Web requests to such applications can be spatially distributed so as to reduce the carbon footprint associated with their execution.
%This study introduces a novel carbon-aware request execution system that encompasses a server provisioning and scheduler, with the primary objective of minimizing the carbon cost of execution. We formulate this system as an optimization problem and construct a prototype for simulate its behavior. Empirical studies reveal that our proposed approach can achieve a substantial reduction in the carbon cost of request execution, with improvements of up to 74\% compared to baseline methods.
%
%There is a crucial need for increased visibility and optimization of carbon efficiency at the datacenter level. 
There has been a significant societal push 
%reduce carbon emissions and 
towards sustainable practices, including in computing.
Modern interactive workloads such as geo-distributed web-services exhibit various spatiotemporal and performance flexibility, enabling the possibility to adapt the location, time, and intensity of processing to align with the availability of renewable and low-carbon energy. An example is a web application hosted across multiple cloud regions, each with varying carbon intensity based on their local electricity mix. 
%By spatially load-balancing web requests, it becomes feasible to reduce applications operational carbon footprint.
Distributed load-balancing enables the exploitation of low-carbon energy through load migration across regions, reducing web applications carbon footprint.
%However, users still face challenges in efficiently optimizing their footprint due to a lack of insights into carbon information.
In this paper, we present \systemName, a carbon-aware scheduling and
provisioning system that primarily minimizes the carbon footprint of distributed web services while also respecting their Service Level Objectives (SLO). 
We formulate \systemName as an multi-objective optimization problem that considers both the variable carbon intensity and latency constraints of the network. Our evaluation reveals the significant potential of \systemName  in achieving substantial reductions in carbon emissions. Compared to baseline methods, \systemName demonstrates improvements of up to 70\% with no latency performance degradation.
\end{abstract}

\begin{CCSXML}
	<ccs2012>
	<concept>
	<concept_id>10010520.10010521.10010537.10003100</concept_id>
	<concept_desc>Computer systems organization~Cloud computing</concept_desc>
	<concept_significance>500</concept_significance>
	</concept>
	<concept>
	<concept_id>10010520.10010521.10010542.10011714</concept_id>
	<concept_desc>Computer systems organization~Special purpose systems</concept_desc>
	<concept_significance>500</concept_significance>
	</concept>
	</ccs2012>
\end{CCSXML}

\ccsdesc[500]{Computer systems organization~Cloud computing}
\ccsdesc[500]{Computer systems organization~Special purpose systems}

\maketitle

\section{Introduction}
%\vspace{1.0cm}
\begin{figure}
	\centering
	\includegraphics[width=\columnwidth]{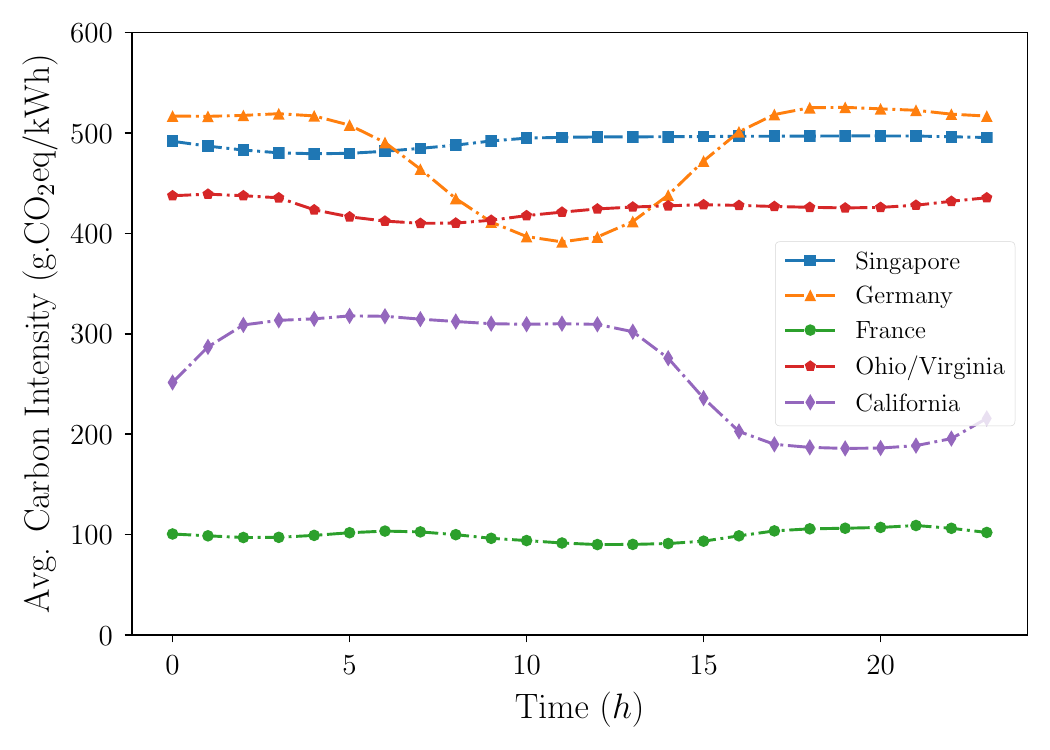}
	\vspace{-0.5cm}
	\caption[]{\textbf{\emph{Grid carbon intensity in 2022 across six distinct cloud regions showing 6$\times$ spatial variations.}}}
	\label{fig:carbon_hourly}
	%\vspace{-0.7cm}
\end{figure}

In recent years, the global focus on sustainability and environmental responsibility has brought renewable energy to the forefront of the discussions on energy systems, leading to an increased focus on reducing the carbon footprint of cloud platforms in both research and industry~\cite{mccallum,ecovisor,chasing,embodied2,embodied3}. 
Although there has been substantial progress in improving efficiency, today's datacenter infrastructures consume around three to five percent of electricity worldwide and in ten years, five times as much \cite{cook2015clicking, jones2018stop, estimate2}.
%estimates indicate that cloud datacenters currently consume $\sim$3\% of global electricity~\cite{estimate} and this percentage is expected to rise to 13\% by 2030~\cite{estimate2} due to the increasing demand for computing and diminishing energy-efficiency improvements. 
These estimations may be lower than reality, as the growth of computing demand has been increasing exponentially for decades~\cite{cacm-article}. %, particularly with the growing popularity of AI workloads. 
%In fact, a recent analysis found that the amount of computation required to train state-of-the-art machine learning models doubled every $\sim$2 years before 2012, following Moore's Law, but has since doubled every $\sim$3.4 months~\cite{ml-compute-demand}, $\sim$7$\times$ times faster. 
%which is triggering serious datacenter efficiency concerns because demand is outpacing supply~\cite{mittal2016survey}, %, which will in turn slow digitization advances down \cite{mittal2016survey}. 
Cloud datacenters have mainly relied on enhancements in energy efficiency, which is unlikely to lead to significant reductions in carbon emissions % alongside agriculture and transportation~\cite{cacm-article}. 
%However, in recent years, the global focus on sustainability and environmental responsibility has brought renewable energy to the forefront of the discussions on energy systems, leading to an increased focus on reducing the carbon footprint of cloud platforms in both research and industry~\cite{mccallum,ecovisor,chasing,embodied2,embodied3}. 
%Major technology companies have recently made public commitments to eliminate their carbon and other greenhouse gas emissions within the next 10-20 years~\cite{amazon-carbon-neutral,facebook-carbon-neutral,vmware-carbon,google-carbon-free,microsoft-carbon-negative}.
%
%
%Unfortunately, relying solely on enhancements in energy efficiency is unlikely to lead to significant reductions in carbon emissions. This is because modern datacenters have already achieved high levels of optimization in energy efficiency. 
as modern datacenters have already achieved high levels of optimization in energy efficiency. 
For instance, the Power Usage Effectiveness (PUE), a measurement of the total operational efficiency of most datacenters, is already near the optimal value of $1.0$. 
More importantly, thesse trends are positioning cloud platforms as one of the largest contributors to global emissions~\cite{cacm-article}. 
Therefore, while energy efficiency improvements are important, they will be insufficient to counterbalance the rising energy consumption from the rapidly growing demand for cloud services.
To effectively reduce carbon emissions, cloud platforms must shift their focus towards low-carbon energy sources. This entails harnessing energy derived from renewable sources such as wind, solar, hydro, nuclear, geothermal, and other sustainable alternatives.
To reduce cloud platforms' carbon emissions, many have suggested leveraging computing workloads' spatial and temporal flexibility, which is often significant, to dynamically shift the location and time of execution to better align with when and where low-carbon energy is available.  Yet, despite the prominence of such simple carbon-aware spatiotemporal workload shifting as an abstract idea, prior work has only quantified its benefits in specific settings such as batch workloads.
Web applications, in particular, serve as an excellent case for exploring the untapped potential of carbon-aware computing. These applications are typically distributed across multiple cloud servers located in different regions worldwide. Traditional approaches reduce latency by forwarding user requests to the geographically closest replica server, reducing load times and offering a better user experience.  On the other hand, different cloud regions have varying carbon costs associated with their electricity sources, leading to different carbon footprints for user requests depending on which replica server services them. Consequently, optimizing the scheduling of user requests with respect to the carbon costs associated with different replicas presents an intriguing opportunity to make web applications more sustainable.

While renewable energy still continues to be unreliable due to its dependence on natural factors, web applications can still benefit from it without sacrificing performance. 
For instance, the inherent fault tolerance achieved through replication and load-balancing mechanisms can safeguard web applications against the intermittency and unpredictability of renewable energy sources~\cite{maji2023bringing}. By ensuring that replicas are spread across diverse regions and backed by different energy sources, these applications can maintain high availability while capitalizing on the potential of cleaner energy.
However, although cloud providers maintain the information about the energy supply powering their servers, it is not readily available at a software level to applications~\cite{ecovisor}. Consequently, resource provisioners and load balancers cannot leverage this information to optimize the carbon efficiency of web workloads. Providing these systems with more visibility into the carbon footprint across datacenters enable the design of heuristics to provision resources and schedule workloads towards replicas with low carbon intensities while respecting applications Service Level Objectives (SLOs).

%This study aims to analyze performance of spatial server provisioning and geo-distributed request scheduling in a distributed web application to optimize on carbon cost of computation The study is divided into two main parts. In the first part, we formalize the problem as an optimization problem. In the second part, we design a prototype that leverages the results of the optimization problem to provision resources and load balance user requests.

As such, we present \systemName, a carbon-aware scheduler and provisioner for distributed web applications. 
We assume a setting where resource provisioning and load balancing ought to happen in concert to minimize emissions while meeting application SLO targets.
This is formulated as a multi-objective optimization problem, addressing both the carbon footprint resulting from server provisioning and the latency caused by load balancing.
%We formulate it as a multi-objective optimization problem that simultaneously considers the carbon footprint due to server provisioning and the latency due to load balancing.
We evaluate the performance of spatial server provisioning and geo-distributed request scheduling for distributed web applications
%with a focus on optimizing the carbon footprint and latency costs of computation. 
%We assume full information of energy data, 
by implementing \systemName as a Kubernetes scheduler and submitting it to a real web workload. 
Our results highlight \systemName's significant potential in achieving considerable reductions in carbon emissions while meeting all latency constrains. In comparison to baseline methods, our approach demonstrates enhancements of up to 70\% without compromising latency performance.
%We release \systemName as an open-source\footnote{\href{https://github.com/carbonfirst/casper}{https://github.com/carbonfirst/casper}} tool
We release \systemName as an open-source tool
that can perform carbon optimizations for their distributed web applications: \href{https://github.com/carbonfirst/casper}{https://github.com/carbonfirst/casper}

%The study is divided into two main components. In the first part, we formulate the problem as an optimization problem, defining the objectives and constraints. In the second part, we develop a prototype that utilizes the outcomes of the optimization problem to effectively allocate resources and balance user requests for improved performance.

\section{Background}
This section provides background on the grid carbon intensity, cloud model, and carbon-aware workload optimizations.

%\subsection{Carbon Efficiency vs Energy Efficiency}
%\noindent\textbf{Carbon Efficiency} 
%In response to the need for mitigating the impact of computing on climate change, researchers have shifted their focus towards optimizing the carbon efficiency of computing~\cite{enabling-socc21, ecovisor}, which measures the amount of work accomplished per unit of carbon emitted. Traditionally, the emphasis has been on enhancing the energy efficiency of computing, quantified as the number of cycles executed per unit of energy consumed, for two main reasons. Firstly, there is a direct incentive to reduce operational costs by improving energy efficiency. Secondly, information regarding the carbon intensity of the electricity powering data centers was either inaccessible or available only at a coarse level of granularity. As a result, energy was treated uniformly across all times and locations, leading to an optimization focus on energy efficiency rather than carbon footprint. However, recent third-party services like electricityMaps and WattTime have closed the information gap, offering real-time data on the carbon intensity of electricity for most locations around the world, facilitating research efforts to optimize the carbon efficiency of computing.

%\subsection{Carbon Intensity}
%\noindent\textbf{Carbon Intensity} 
\subsubsection*{Carbon Intensity} The electric grid relies on a combination of generation sources to meet the demand for electricity, and include fossil fuel-based generators using coal or natural gas, low-carbon sources like hydro, wind, and solar, as well as non-carbon sources such as nuclear. Since electricity demand 
fluctuates throughout the day and follows diurnal patterns, the mix of generation sources and their relative proportions also vary over time. It is worth noting that renewable sources --- such as wind and solar --- are intermittent, which further impacts the overall generation mix. The carbon intensity (CI) of electricity supply, measured in grams of CO2 equivalent per watt or g$\cdot$CO$_2$eq/kWh, represents the average weighted carbon intensity of the generation sources used at any given moment. As fossil-based sources have high, and renewable sources have low or zero carbon weights, the average CI depends on the proportion of each source in the overall generation mix. 
Figure \ref{fig:carbon_hourly} illustrates the average carbon intensity of the grid electricity over the 2022 period for six different geographical regions, revealing significant variations across locations.
On the vertical axis, the carbon intensity exhibits spatial variations between regions, while the horizontal axis presents temporal variations within regions. %the day due the the daily changes in the generation mix. 
% As shown, Ontario has the lowest carbon-intensity due to its use of nuclear power, while Uruguay has a slightly higher carbon-intensity due to its use of hydroelectricity. California has the highest carbon-intensity due to the use of fossil fuels, but also the highest variability due to its high solar penetration.
As shown, France has the lowest carbon-intensity due to its reliance on nuclear power, while Germany and Singapore have the highest intensities due to their reliance on fossil fuels.
However, regions like California and Germany have higher temporal variability due to increasing penetration of renewables.
%and an even higher 6-8$\times$ variation in Germany and California for the same period.
These variations imply that the carbon footprint of a job can vary by up to 40\% depending on whether it is executed during a high or low carbon-intensity period. %In addition, they suggest that the same job executed in different cloud regions results 6-8$\times$ variation in different emissions, underscoring the potential for techniques that intelligently schedule workloads on clusters based on the current and projected carbon intensity of grid electricity. 
Moreover, they indicate that executing the same job in different cloud regions can lead to a 6-8x variation in emissions. This underscores the potential for techniques that strategically schedule workloads on clusters based on the current and projected carbon intensity of grid electricity.
While cluster managers can leverage temporal variations by aligning job execution with low carbon periods~\cite{ecovisor, wait-awhile}, we focus on exploiting spatial characteristics that involves distributing workloads across regions with both low carbon intensity and sufficient latency performance.
%While cluster managers can leverage the temporal variations by aligning jobs execution with low carbon periods, our current work primarily focuses on exploiting the spatial characteristics by geographically distributing workloads across regions with low carbon intensity and enough latency performance.
%
%While geographically-federated clusters can leverage these spatial variations by scheduling incoming jobs on cluster resources in regions with lower carbon intensity,  temporal variations within a single cloud region. Addressing spatial batch scheduling across geographically federated clusters is left as a topic for future research.
Finally, our work concentrates on scheduling techniques aimed at reducing scope 2 emissions as defined by the GHG (Greenhouse Gas) protocol~\cite{ghg}, in which the majority of operational emissions are attributed to energy consumption (including scope 1). We do not consider embodied emissions (scope 3).
%\noindent\textbf{Spatiotemporal Shifting} 
\subsubsection*{Workload Flexibility}
%\noindent\textbf{Workload Flexibility} 
%\subsubsection*{Workload Flexibility} 
The potential for a job to reduce emissions is a function of its type -- batch or interactive --, memory state, and the network latency and bandwidth across locations.
%Particularly for spatial shifting
%Particularly, a job's overhead is a function of 
Additionally, there may be regulatory constraints, such as HIPPA~\cite{assistance2003summary} and GDPR~\cite{gdpr}, that prevent spatially shifting a job outside of a specific country, region or jurisdiction. 
%We consider the impact of such regulatory constraints in \S\ref{sec:spatial}. 
%As mentioned, the variations in carbon intensity of electricity supply have a direct impact on the emissions generated when computing across different locations. 
\begin{figure}[!ht]
	\centering
	\includegraphics[width=\columnwidth]{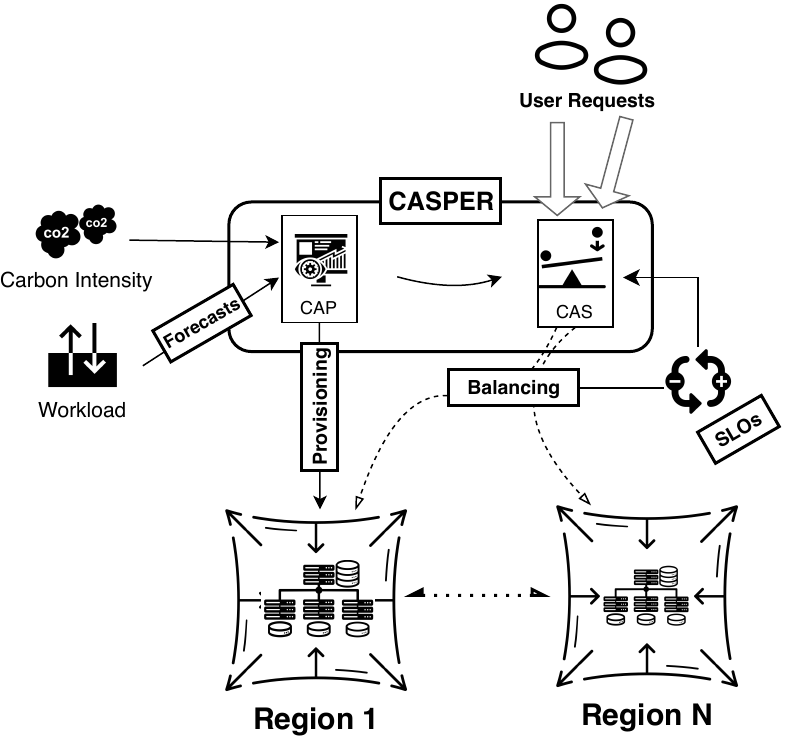}
	\vspace{-0.3cm}
	\caption[]{\textbf{\textit{\systemName: CAP and CAS provision and coordinate user workloads.}}}
	\label{fig:system}
\end{figure}
While batch jobs such as AI and machine learning often have flexible completion times and can accommodate temporal variations, interactive workloads have strict low-latency requirements and limited temporal flexibility. 
This is especially true in web-services environments where requests pass through multiple microservices before a response is produced. 
%Considering that many services are highly optimized for latency performance, slight variations in latency under a target are unlikely to affect the user experiment and allows for the spatial flexibility dimension can be leveraged to optimize carbon emissions. 
%Considering that users requests already accustomed to some requests experiencing slightly higher latency, such spatial flexibility dimension can be leveraged to optimize carbon emissions. 
%By strategically scheduling workload processing at locations with lower carbon intensity, carbon optimization can be achieved without significantly compromising user experience.
For instance, load balancing tools enable modern workloads with the ability to shift their execution location to minimize latency and improve service availability~\cite{souza2018hybrid}. 
%align with renewable energy sources and that are cost-effective.  
These techniques work mostly with workloads that have lightweight memory states and which do not need to transfer data around locations.
In this study, we consider lightweight web-requests -- specifically HTTP requests --, that can be seamlessly processed across various locations. These requests have latency requirements that need to be limited within a maximum threshold.
Given that numerous services are highly optimized for latency, minor deviations within a specified target are unlikely to impact the overall user experience, and can enable the exercise of spatial shifting to optimize for carbon.

\section{System Design and Implementation}
This section outlines the design and architecture of \systemName, along with its key components. % required for its carbon-awareness.
%We also introduce a load-balancer that optimizes the workload performance by redirecting requests across servers.

\subsection{Architecture}

\systemName is designed as a modular system that can be integrated into any existing distributed resource manager.  Figure \ref{fig:system} illustrates the overall system architecture, highlighting its two main components: the Carbon-Aware Provisioner (CAP) and the Carbon-Aware Scheduler (CAS). \systemName includes various components for interfacing with interactive jobs, such as the resource manager, monitoring, and the carbon-aware load-balancing and scheduling policies.
%
%\noindent
\subsubsection*{\textbf{Carbon-Aware Provisioner.}} 
%\subsubsection*{Carbon-Aware Provisioner. } 
CAP acts as an intelligent 
%decision-making 
provisioner that analyzes the inter-regional network latency, the region's (variable) carbon intensity, and the expected application's workload. Besides reducing carbon, CAP provides operators with an important estimator: the optimal number of servers needed in each region such that the expected workload is correctly handled for each time period and lowest carbon intensity. %, and (ii) the optimal distribution of requests among the regions.
%The proposed architecture has two pivotal components: The Carbon-Aware Provisioner (CAP) and the Scheduling system. CAP functions as an intelligent decision-making module, which processes data on inter-region latency, region-wise carbon intensity and expected workload for the hour, and outputs two crucial metrics: The region-wise count of servers required to efficiently serve the next hour's requests and the optimal distribution of requests between regions.
%\subsection{Carbon Aware Provisioner}
\begin{table}[!ht]
	\centering
	\resizebox{0.8\columnwidth}{!}{%
		\begin{tabular}{@{}cl@{}}
			\toprule
			\multicolumn{1}{l}{\textbf{Parameter}} & \textbf{Description}                           \\ \midrule
			$x_{ij}$                               & Requests redirected from region $i$ to region $j$ \\ \midrule 
			$\bar{x_j}$ & Requests not sent to region $j$ \\ \midrule 
			$s_j$                                  & Number of servers in region $j$ \\ \midrule
			$n$                                    & Number of regions $\mathcal{R}$                              \\ \midrule
			$I_j$                                  & Carbon intensity in region $j$                 \\ \midrule
			$\alpha$ & \begin{tabular}[c]{@{}l@{}}Normalized weight for the carbon intensity \\ (in relation to number of servers $s_j$)\end{tabular} \\ \midrule
			$\lambda_i$                            & Incoming request rate at region $i$            \\ \midrule
			$\ell_{ij}$                            & Expected latency from region $i$ to $j$        \\ \midrule
			$c_j$                                  & Resource capacity of region $j$ (in \# of requests)               \\ \midrule
			$L_i$                                  & Maximum tolerated latency for a request        \\ \midrule
			$K$                                    & Maximum number of servers across all locations \\ \midrule
			$t_{j}$                                & Number of requests submitted to region $j$  \\ \bottomrule
		\end{tabular}%
	}
	\caption{\textbf{\textit{List of parameters used by CAP.}}}
	\label{tab:equations}
	%\vspace{-0.99cm}
	%\vspace{-0.6cm}
\end{table}
%To estimate the number of servers for a region, we devise a mixed-integer multi-objective optimization problem that minimizes the total carbon cost of executing requests and the cumulative number of servers sprung up across all regions. 
%A single objective problem where only carbon intensity is being minimized could be devised. However, an important downside is that such model would not penalize the provisioner for allocating more servers than necessary in any given region. 
%
%While minimizing the carbon intensity, our formulation ensures that the provisioner deploys no more servers than required to service the expected workload. 

This intuition leads us to formalize this provisioning problem as a \textit{multi-objective formulation}: %instead of a single-objective optimization problem.
%To address the challenges of carbon-aware provisioning, we have developed a mixed-integer multi-objective optimization problem. 
%\vspace{-1cm}
\begin{subequations}
    \begin{align}
    \underset{x}{\min} \quad & \alpha \sum_j I_j\sum_i x_{ij} + (1-\alpha) \sum_j s_j\\
    \text{s.t.}  \quad & \sum_i x_{ij} \leq s_jc_j \\
    & \sum_j s_j \leq K \\
    & x_{ij}\left(\ell_{ij}-L\right) \leq 0 \\
    & \sum_{i,j} x_{ij} = \mathbb{E}[\lambda_i], \forall i,j \\
    & \alpha \in [0,1] \\
    & \bar{x}_js_j=0 \\
    & x_{ij},s_j \in\mathbb{Z}_{\geq 0}
    \end{align}
    \label{provEq}
    %\vspace{-0.2cm}
\end{subequations}
%\begin{equation*}
%\begin{aligned}
%n &:= \text{Number of regions}\\
%\alpha &:= \text{Multi-objective optimization weight}\\
%I_j &:= \text{Carbon intensity in region $j$}\\
%\lambda_i &:= \text{Incoming request rate at region $i$}\\
%\ell_{ij} &:= \text{Latency from region $i$ to $j$}  \\
%c_j &:= \text{Capacity for each server in region } j \\
%L_{i} &:= \text{Maximum tolerated latency for request redirection} \\
%K &:= \text{Maximum number of servers } \\
%\end{aligned}
%\end{equation*}
%
%\subsubsection{CAP Formulation}
%\label{sec:capm}
%
%\vspace{-0.1em}

We present CAP's formulation in Equation \ref{provEq}, and Table \ref{tab:equations} describes all of its parameters. 
All indices $i, j$ represent the set of available regions $\mathcal{R}$ for resource allocation, request processing and redirection capacities. We let $\bar{x_j}\in\{\,0,1\,\}$ be the variable that represents requests that are not sent to a region $j$, i.e., $\sum_j x_{ij}=0$.
%This constraint essentially implies that if region $j$ is not expected to receive any requests, then no servers should be deployed there.
This constraint effectively means that if it is anticipated that region $j$ won't receive any requests, there should be no server allocation in that region.
Moreover, $x_{ij}$ represents the optimal count of requests from region $i$ that is redirected to region $j$, while $s_j$ is the number of servers provisioned in region $j$ to handle incoming requests.
Eq. \ref{provEq}a aims to minimize both the total carbon footprint of executing requests (Eq. \ref{provEq}b) and the cumulative number of servers $s_j$ across all regions (Eq. \ref{provEq}c) such that the minimum latency target is guaranteed (Eq. \ref{provEq}d). 
%Alternatively, we could approach it as a single-objective problem by solely minimizing emissions through the carbon intensity. However, this approach has a drawback in that it cannot penalize the provisioner for deploying more servers than necessary in a specific region.
%To ensure that the CAP formulation deploys the minimum required number of servers while minimizing carbon emissions, we have chosen to formalize the problem as a multi-objective optimization problem rather than a single-objective optimization problem.
%\subsection{Policy Model}
Finally, the following assumptions are made.
First, the problem is defined within the scope of minimizing carbon emissions while simultaneously adhering to application latency constraints.
%the scope of the problem is limited to minimizing carbon emissions while respecting applications latency constraints. 
Second, since we consider cloud datacenters, we ignore issues regarding resource limits, although we do include a maximum amounr of servers (Eq. \ref{provEq}c) that \systemName can provision.
We also assume the load-balancers communication latency across regions (as seen in Figure \ref{fig:system}) is negligible when compared to the requests' average service (processing) times.
Finally, the provisioner uses forecasts for carbon intensity~\cite{maji2022carboncast} and hourly workload request rates that are expressed in terms of expected arrivals in region $i$ (Eq. \ref{provEq}e).
%\begin{itemize}
%    \item The scope of the problem is limited to minimizing carbon emissions while respecting applications latency constraints. Since we consider cloud datacenters, we ignore issues such as capacity planning or resource limits.
%    \item We assume low latency communication across load-balancers in different regions (as seen in Figure \ref{fig:system}).
%    \item The provisioner uses forecasts for carbon intensity and hourly workload request rates.
%\end{itemize}
%

\subsection{Carbon Aware Scheduler}
\vspace{-3ex}
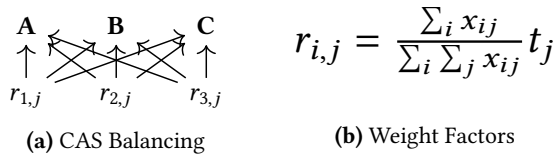
\begin{figure}[H]
    %\centering
%    ~ 
    \begin{subfigure}[t]{0.45\columnwidth}
        %\centering
\[\begin{tikzcd}[column sep=3ex,row sep=3ex]
	{\textbf A} & {\textbf B} & {\textbf C} \\
	{r_{1,j}} & {r_{2,j}} & {r_{3,j}}
	\arrow[from=2-2, to=1-1]
	\arrow[from=2-2, to=1-2]
	\arrow[from=2-2, to=1-3]
	\arrow[from=2-1, to=1-1]
	\arrow[from=2-1, to=1-2]
	\arrow[from=2-1, to=1-3]
	\arrow[from=2-3, to=1-1]
	\arrow[from=2-3, to=1-2]
	\arrow[from=2-3, to=1-3]
\end{tikzcd}\]
\vspace{-1.7ex}
        \caption{CAS Balancing}
        \label{fig:distrbuted_scheduler}
    \end{subfigure}
        \begin{subfigure}[t]{0.5\columnwidth}
%        \centering
        %\begin{subequations}
        ~
\begin{myequation}%
%\begin{align*}
r_{i,j} = \frac{\sum_i x_{ij}}{\sum_i \sum_j x_{ij}} t_{j}
%\end{align*}
\end{myequation}
\label{casEq}
%\end{subequations}
\vspace{1.0ex}
        \caption{Weight Factors}
        \label{fig:central_scheduler}
    \end{subfigure}%
    \caption{\textbf{\textit{Illustration of CAS and weight calculation.}} }
    \label{fig:scheduling_approach}
    \vspace{-0.4cm}
\end{figure}

Figure \ref{fig:scheduling_approach} shows how the Carbon Aware Scheduler (CAS) 
%applies the request distribution from CAP and estimate how to 
distributes requests between regions $x_{ij}$, which denotes the load of requests that need to be redirected from region $i$ to $j$.
%
%\subsubsection{Parameters}
%
%\begin{equation*}
%\begin{aligned}
%r_{j} &:= \text{Number of requests received by server backend } \\
%&\quad\text{in region j from CAS} \\
%t_{j} &:= \text{Number of requests received at region j from client } \\ 
%\end{aligned}
%\end{equation*}
%
%Each server uses a $r_j$ vector that calculates a weighted factor for each pair of regions $x_{i,j}$ as shown Eq. \ref{fig:scheduling_approach}(b).
%
%\begin{subequations}
%\begin{align}
%r_{j} = \frac{\sum_i x_{ij}}{\sum_i \sum_j x_{ij}} t_{j}.
%\end{align}
%\label{casEq}
%\end{subequations}
%
It uses a vector as shown by Equation \ref{fig:scheduling_approach}(b) to model each region $r$'s weight, following the timely estimates obtained from CAP (e.g., hourly). We implement CAS as a load-balancer module in \systemName, whereas local incoming requests are redistributed across all regions according to their proportional weights. 
%This is important because in this way, sudden unexpected workload spikes not considered in the CAP's optimization can properly be handled.
More importantly, CAS ensures that unforeseen workload events --- e.g., load spikes, not accounted for in CAP's optimization --- are effectively handled as best as possible.
%Finally, the CAS proportional weights are updated following CAP's updates (e.g., hourly).

\subsection{Implementation}

\systemName is implemented as a Kubernetes (K8S) scheduler with $s < K$ workers, each representing a cloud region. Each deployment comprises of a single K8S pod that runs the application.
A prototype has been developed to emulate the operations of a Wikipedia-like service across six distinct AWS regions, as detailed in Table \ref{tab:regions}. 
These regions are selected as the closest regions to the original Wikimedia servers~\cite{wikimedia_infra}.

%\noindent\textbf{CAP.} CAP has been developed using Python3 and is run at an hourly granularity. The optimization problem is solved using the PuLP library, a python interface to the Coin-or branch and cut (CBC) solver. The region wise server deployment array obtained as an output of CAP is used to vertically scale the size of each Kubernetes deployment using metrics server. On performing load testing, it has been observed that each deployment/pod can handle a maximum of 1000 requests/second at full capacity. Assuming that each region can have a maximum of 100 servers, each server is analogous to 0.01 Pod CPU. Therefore, at startup, each pod is allotted 0.01 CPU, and is thereafter allotted more or less based on CAP output. The vertical scaling has been performed purely on CPU cores and not on memory because we have observed that CPU is the sole bottleneck. 
%Additionally, CAP computes the optimal request distribution matrix, which is forwarded to CAS.

%\noindent
\subsubsection*{\textbf{CAP}} \systemName's provisioner is developed using Python, while the optimizations are solved using the PuLP library~\cite{mitchell2011pulp}, an interface to the Coin-or branch and cut (CBC) solver~\cite{saltzman2002coin}. The region wise server deployment array obtained as an output of CAP is used to scale the size of each regions using server collected metrics. Additionally, CAP computes the optimal request distribution matrix, which is forwarded to CAS. % vertically
%On performing load testing, it has been observed that each deployment/pod can handle a maximum of 1000 requests/second at full capacity. Assuming that each region can have a maximum of 100 servers, each server is analogous to 0.01 Pod CPU. Therefore, at startup, each pod is allotted 0.01 CPU, and is thereafter allotted more or less based on CAP output. The vertical scaling has been performed purely on CPU cores and not on memory because we have observed that CPU is the sole bottleneck. 

%\noindent\textbf{CAS.} CAS system utilizes a set of load balancers, with one deployed in each region, to direct incoming requests to the appropriate regions based on weights derived from the CAP's optimal request distribution matrix. Traefik~\cite{traefik} is used to establish the load balancer layer for the cluster setup. It creates an HTTP proxy for every region from which requests are received. Each traefik proxies can dynamically route the traffic to one of the Kiwix backends based on weights updated hourly by CAP. Using Traefik has multiple advantages over other open-source load-balancers available:

%\noindent
\subsubsection*{\textbf{CAS}} The scheduler coordinates a set of load balancers, one per-region, to implement its logic. It timely forwards incoming requests to the appropriate regions following the weights derived in the CAP's optimal request distribution matrix. Traefik~\cite{traefik} is used to establish the cluster's load balancer layer, creating a HTTP proxy for every region to receive and forward requests by routing the traffic to one of the corresponding backend regions based on the hourly weights calculated by CAP. 

%Using Traefik has multiple advantages over other open-source load-balancers available:
%\begin{itemize}
%    \item A traefik service is present in every region. This design aligns with the assumption that the latency between the client and the load balancer is negligible.
%    \item Traefik client allows for seamless updates of load-balancing weights throughout the simulation.
%    \item Traefik can be easily integrated with Prometheus enabling the collection of valuable metrics during the simulation process.
%\end{itemize}

\section{Evaluation}

\begin{figure*}[!ht]
    %\centering
    \begin{tabular}{cccc}
    \includegraphics[width=0.2215\textwidth]{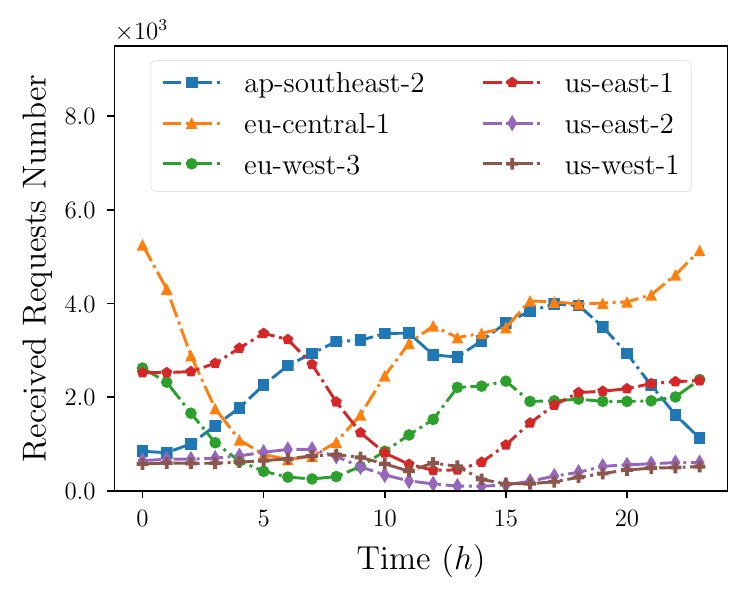} &
    \includegraphics[width=0.2215\textwidth]{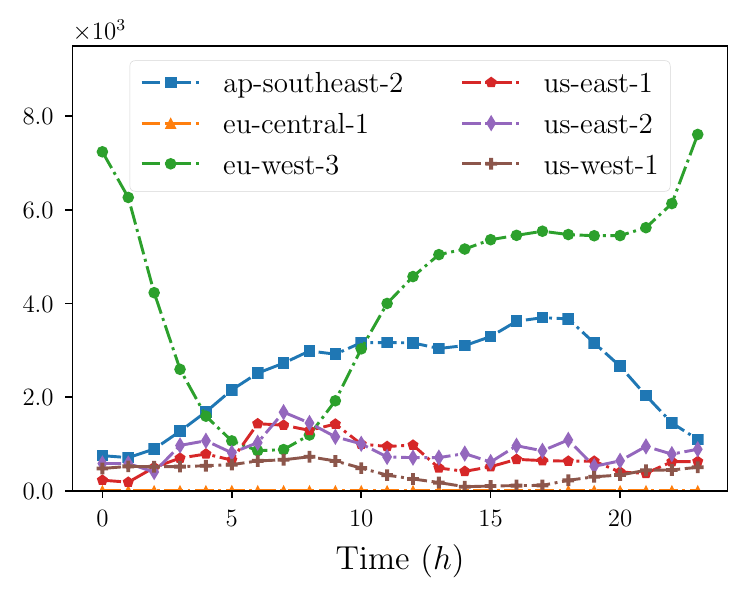} &
    \includegraphics[width=0.2215\textwidth]{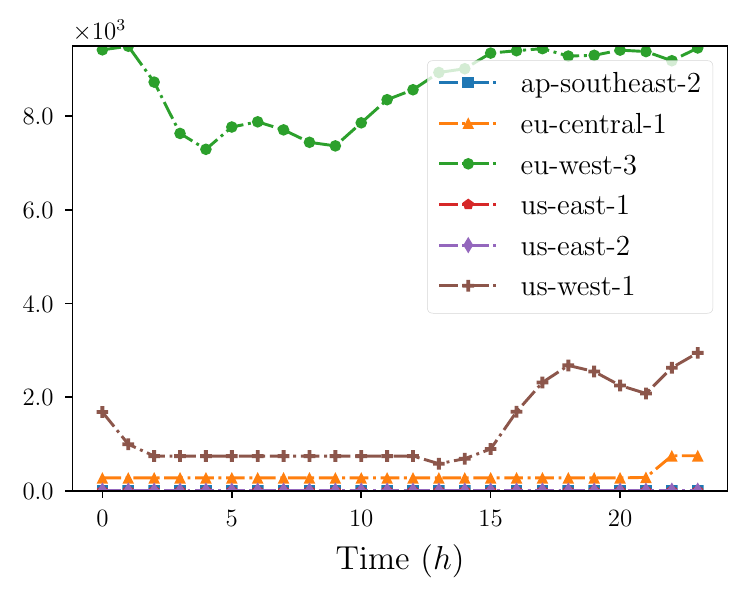} &
    \includegraphics[width=0.273\textwidth]{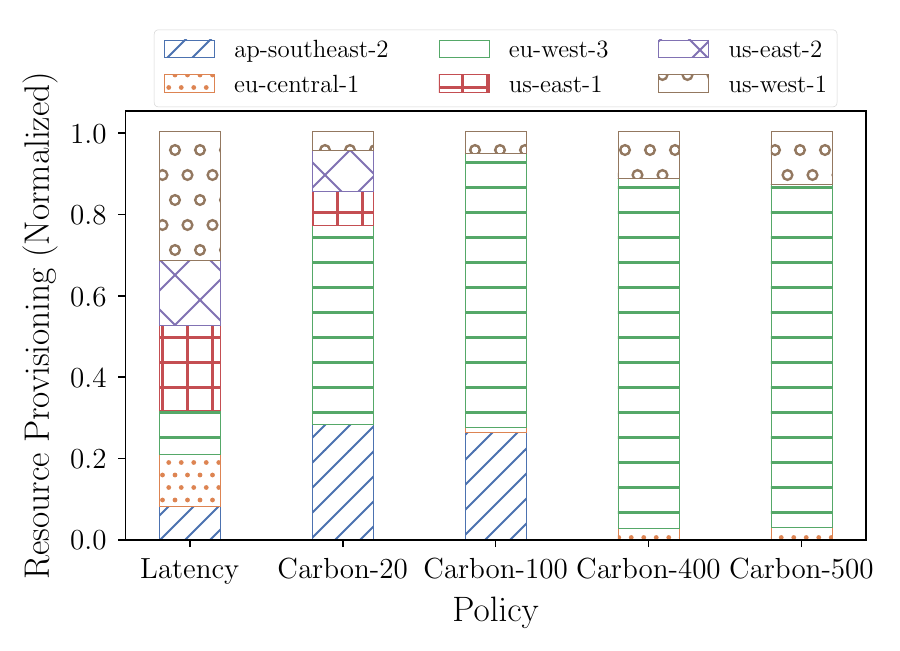} \\
    (a) Latency &  (b)  Carbon-20 & (c) Carbon-500 & (d) Resource Utilization
    \end{tabular}
    \vspace{-0.4cm}
    \caption{\textbf{\emph{Redirection rate (a, b and c) and resource provisioning per-region and policies (d): Provisioning tends to increase in greener nearby regions.}}}
    \vspace{-0.4cm}
    \label{fig:replay_requests}
\end{figure*}

In this section, we first discuss the real-world application, workload, carbon, and network traces that are utilized to evaluate \systemName.
Then, we discuss the policies briefly introduced in the previous section. 
Finally, we demonstrate and quantify the trade-offs between carbon savings and latency performance for various targets.

\begin{table}[]
\centering
\resizebox{0.7\columnwidth}{!}{%
\begin{tabular}{@{}ll@{}}
\toprule
\textbf{Geographical Region} & \textbf{AWS Region} \\ \midrule
California                   & US-West-1           \\ \midrule
Virginia                     & US-East-1           \\ \midrule
Ohio                         & US-East-2           \\ \midrule
Germany                    & EU-Central-1        \\ \midrule
France                        & EU-West-3           \\ \midrule
Singapore                    & AP-SouthEast-2      \\ \bottomrule
\end{tabular}%
}
\caption{\textbf{\textit{AWS Regions used in the evaluation.}}}
\label{tab:regions}
\vspace{-0.8cm}
\end{table}

\subsection{Setup}

\begin{comment}
\begin{table}[H]
    \centering
    \begin{tabular}{|c|}
        \hline
        ap-southeast-2 (Singapore) \\
        \hline
        eu-west-3 (Paris) \\
        \hline
        eu-central-1 (Frankfurt) \\
        \hline
        us-east-1 (N.Virginia) \\
        \hline
        us-east-2 (Ohio) \\
        \hline
        us-west-1 (N.California) \\
        \hline
          \caption{Regions used in the simulation.}
    \end{tabular}
    \label{tab:region_data}
\end{table} 
\end{comment}

%\noindent\textbf{Infrastructure. } 
\subsubsection*{Infrastructure. } \systemName runs on Ubuntu Linux 20.04, and it consists of a control plane and worker nodes. The cluster compromises 16 servers with 16-cores Intel Xeon processors and 32GB of memory. Each node runs a Kubernetes deployment representing one region. For intra-cluster communication, an overlay network is created using Flannel~\cite{flannel}. %, which provides communication between containers across nodes.

%\noindent\textbf{Application. } 
\subsubsection*{Application} To evaluate \systemName, we deploy Kiwix~\cite{kiwix}, a platform to host and distribute compressed versions of the Wikipedia~\cite{openzim}. %, with complete built-in full-text search. 
Specifically, we load Kiwix with the pre-built version of the German Wikipedia from May 2023, which comprises a total of 32 GB of content~\cite{wikidata}.
Requests are directed through the CAS load balancer, which interconnects all nodes in the cluster via a HTTP port.
%Requests are sent through the CAS load-balancer that interconnects all nodes in the cluster through a network-wide accessible HTTP port.
%We choose to run Kiwix, which is specifically designed to emulate a Wikipedia server using a custom Zim file.
%Since we are only interested in the intra-network latency and emissions across regions, only the Kiwix frontend page is accessed.

%Others include $traefik\_service\_request\_duration\_seconds\_sum$ and $traefik\_service\_request\_duration\_seconds\_count$, which are employed to compute the average request duration at each traefik endpoint. Since all six regions are deployed within a local cluster, the average request duration is indistinguishable across regions. Consequently, this metric is combined with AWS inter-region latency data obtained from Cloudping to represent the overall latency of request execution.

%\subsection{Energy and Workload Datasets}

%\textbf{1. Carbon Intensity}.
%\label{sec:datasets}
%Where do the request data set come from?
%\noindent
\subsubsection*{\textbf{Carbon Intensity. }} Figure \ref{fig:carbon_hourly} illustrates the carbon intensity data for all the aforementioned geographical regions (Table \ref{tab:regions}) at an hourly granularity. This data has been collected from Electricity Maps~\cite{electricity_map} for 2022.

%\textbf{2. Wikipedia Request Rate}.
%\noindent
\subsubsection*{\textbf{Workload and Network Traces. }} We use the Wikimedia's dataset~\cite{wikimediagrafana} covering six datacenters across the USA, Europe, and Asia. For each region, the dataset includes the request rates (requests per-second) and datacenter hourly utilization covering 2022. Since two of the AWS regions do not match those from Wikimedia's -- i.e., Netherlandas and Texas --, we select the two closest AWS regions i.e., Germany and Ohio (Table \ref{tab:regions}).
Average latency data (in milliseconds) across all AWS regions are obtained from Cloudping~\cite{cloudping} for 2022.

\subsubsection*{\textbf{Telemetry. }} Each region's load-balancer exports their service-level metrics, specifically the total count of HTTP requests served by each endpoint and their associated service time. To calculate the carbon cost of request execution, this metric is multiplied by the region's current hour's carbon intensity.

\subsubsection*{\textbf{Policies}} We conduct a comprehensive evaluation of \systemName throughout the entire year of 2022. The parameters for CAP are set as follows: $n = 6$ (representing the AWS regions), $\alpha$ = 0.5 (equal weights to both carbon and latency costs), $c_j = 100$ (one server can handle up to 1k requests), and $K = 500$ (global maximum number of servers). We also introduce several variations in the values of $L_i$ (see below), which establish the maximum acceptable latency for each request.
CAP runs at the beginning of every hour to determine the provisioning of servers at each location. The CAS weights are then calculated based on the output of CAP. To assess the system's performance, we execute a real workload simulation spanning 24 hours. Metrics are collected at 10-minute intervals and aggregated at the end of each hour. We conduct evaluations using the following policies:

%We evaluate \systemName across all days of the 2022 year. CAP parameter settings are $n$=6 (AWS regions), $\alpha$=0.5, each location $c_j$=10, maximum amount of servers K=100. 
%We vary the values for $L_i$, which sets the maximum tolerated latency for every request. 
%CAP runs at the start of every hour, and servers are provisioned at each location accordingly. CAS weights are computed on the basis of CAP's output. 
%A real workload run of 24 hour is executed, and metrics are collected at intervals of 10 minutes, followed by aggregation at the end of each hour.  %Exponential and binomial distributions have been experimented with for the real workload generation. 
%In particular, the following settings are evaluated:

\begin{enumerate}[leftmargin=*]
\item \textbf{Latency.} This simulation serves as the baseline scenario without any carbon optimization, where requests are solely served based on the lowest latency, i.e., locally in the originating region, without any load balancing. %This configuration strictly aligns with Mediawiki's operational approach, as it adheres to their stringent latency performance requirements.
%Figures 2 and Figure 3 illustrate the count of requests originating from each region and served by each region, respectively. 
%Since the replay simulation executes all requests in the region of origin, both figures exhibit the same distribution. 

%Figure 5 presents a plot of the average per-request latency at each region, which remains constant across all hours, equivalent to the intra-region latency for each region. The carbon cost of request execution in each region is determined by scaling the request count with the carbon cost associated with that region.  

%\begin{figure}[H]
%    \centering
%    \includegraphics[width=0.4\textwidth]{images/replay_requests_to.png}
%    \caption{Replay: Count of requests going to each region's backend}
%    \label{fig:replay_requests}
%\end{figure}

%\begin{figure}[H]
%    \centering
%    \includegraphics[width=0.4\textwidth]{images/replay_carbon_intensity.png}
%    \caption{Replay: Carbon Intensity per region}
%    \label{fig:replay_requests}
%\end{figure}

%\begin{figure}[H]
%    \centering
%    \includegraphics[width=0.4\textwidth]{images/replay_latency.png}
%    \caption{Replay: Avg per request latency per region}
%    \label{fig:replay_requests}
%\end{figure}

\item \textbf{Carbon-L Policies. } These runs focus on carbon optimization with various latency $L$ threshold guarantees, ranging from 20 to 500ms. 
%It is important to note that 
This approach involves a trade-off in terms of performance, as requests can be redirected as long as the latency requirements remain below $L$ ms.
%As such, there is a performance trade-off because requests can be redirected as long as the latency of redirection is less than $L$ ms.
%The scenario with a latency threshold of $L = 500 ms$ is the highest across all possible inter-region latencies. In this case, the carbon cost is minimal because it allows unrestricted redirections across all regions, so as long as each region remains within their maximum request capacity.
\end{enumerate}
%\subsection{Results}

These implementations strictly follow Mediawiki's operations, in particular the Latency policy that adheres to their stringent latency requirements~\cite{wikimedia_infra}. Among the carbon-aware policies, we set one with threshold of $L = 500$ ms as it represents the most flexible response time across all regions. In this particular setting, the carbon cost of execution is minimized by irrestrictive redirections %within their resource capacity.
%and represent cases where the carbon cost of execution is reduced
that can reach very distant, lower carbon regions capable of accommodating the redirected requests. 

%\noindent
\subsubsection*{\textbf{Workload Generation}}
%Workload generation was implemented for the performance evaluation of the prototype. 
%The workload data for CAP optimization was utilized to generate the workload for the prototype. 
A sample of the workload is represented in Figure \ref{fig:replay_requests}(a), with incoming requests in all regions.
Each hour is divided into timesteps, and the request rate for each timestep is selected from a set of values that follow an exponential distribution. Parameters to generate the distributions are selected such that the upper limit of the generated values is approximately 1.5$\times$ the request rate for the hour. %, as this can be handled by the Kubernetes pods without dropping any requests. 

\subsection{Results}

\subsubsection*{\textbf{Effects on Request Redirection}}

Figures \ref{fig:replay_requests}(a)-(c) present workload redirection results for \textit{Latency}, \textit{Carbon-20}, and \textit{Carbon-500}. As shown in Figure \ref{fig:carbon_hourly}, Zone "eu-west-3" (France) has the lowest carbon intensity, followed by "us-west-1" (California), "us-east-[1,2]" (Ohio/Virginia), "eu-central-1" (Germany), and "ap-southeast-2" (Singapore).
Figure \ref{fig:replay_requests}(a) simply shows the original workload, where no redirection happens.
Notably in Figure \ref{fig:replay_requests}(b), due to close proximity and low carbon intensity, \systemName redirects as many requests as possible from Germany towards France. And due to the latency constrains (20ms), Ohio and Virginia cannot induce savings.
This behavior is more evident in \ref{fig:replay_requests}(c): As the latency constraint is relaxed (500ms), France and California receive as many requests as possible from all regions. %"eu-west-3", "us-west-1", and few to "eu-central-1". %to reduce total carbon cost of execution.
%At moments (see Figure \ref{fig:carbon_hourly}), specially when renewable availability increases during the day, these regions exhibit considerably lower carbon costs.
However, eu-west-3 reaches capacity at various moments, triggering \systemName to forward load to California (us-west-1).
\subsubsection*{\textbf{Effects on Resource Provisioning}}

%Figure \ref{fig:rsrc_util} presents results for resource provisioning across the six AWS regions.
%Latency has the original provisioning as no requests are redirected. As response time constraints increase, \systemName starts re-provisioning German ("eu-central-1") servers in France because of its reduced carbon and latency costs.
%With Carbon-100, most of Ohio and Virginia loads are entirely forwarded to France, as the 100ms latency can satisfy the response time.
%Notably, incoming requests originated in Singapore can only reach greener locations through Carbon-400. This is due to the end-to-end latency from Singapore to any other location being larger than 100ms.
%However, Carbon-400 tops the capacity in France, in which case redirections start going through California, and few portions through Germany due to latency requirements.

\begin{figure}[!ht]
    \centering
    \begin{tabular}{cccc}
    \includegraphics[width=0.48\columnwidth]{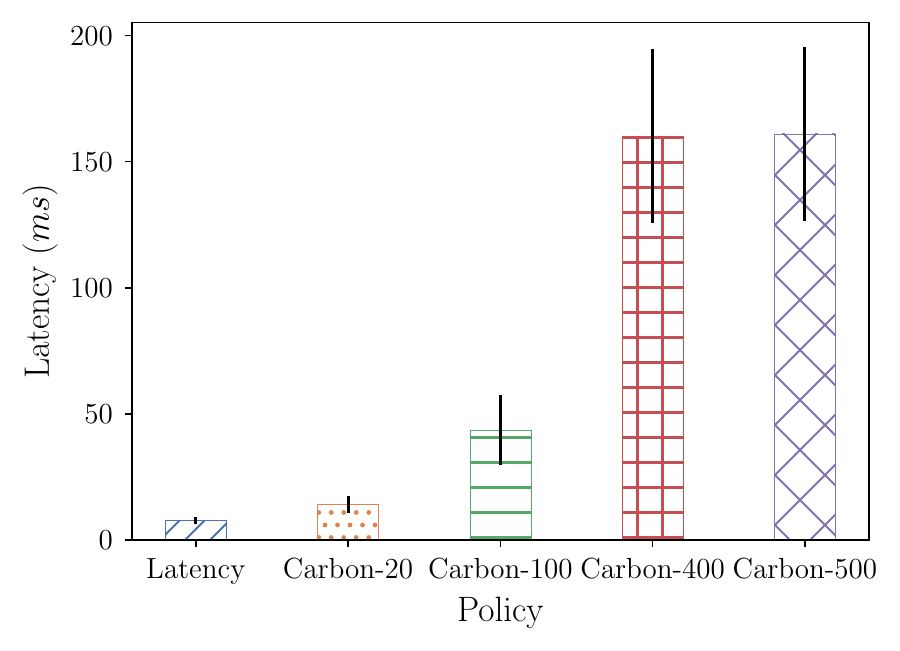} &
    \includegraphics[width=0.48\columnwidth]{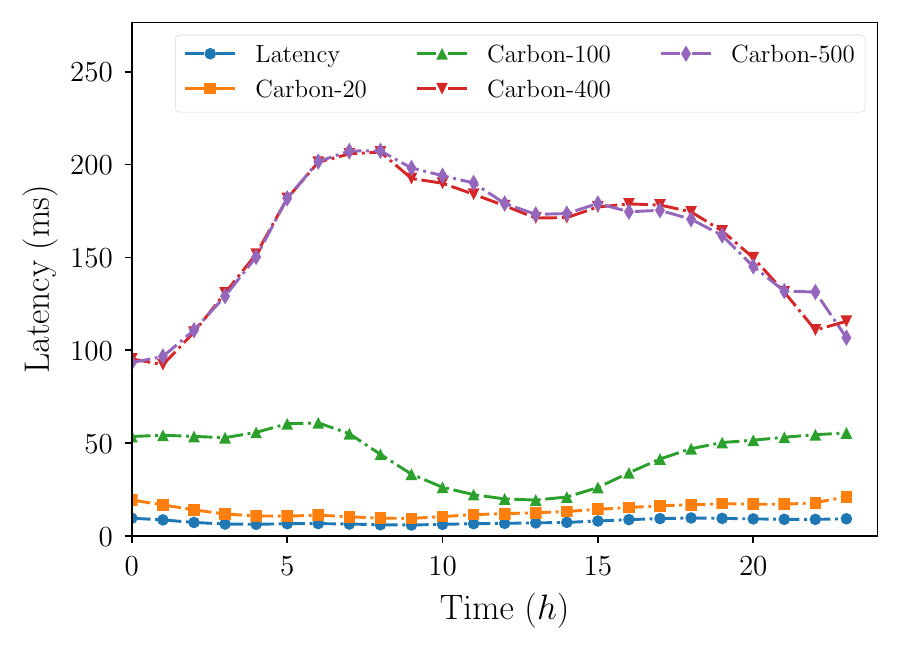} & \\
    (a) Average Latency &  (b) Hourly Latency \\
    \includegraphics[width=0.47\columnwidth]{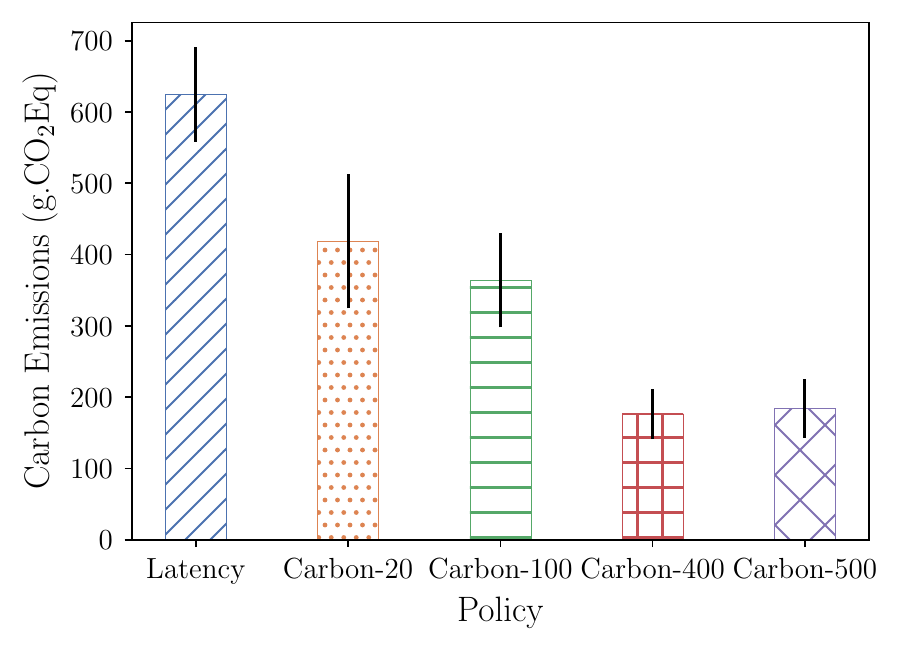}  &
    \includegraphics[width=0.47\columnwidth]{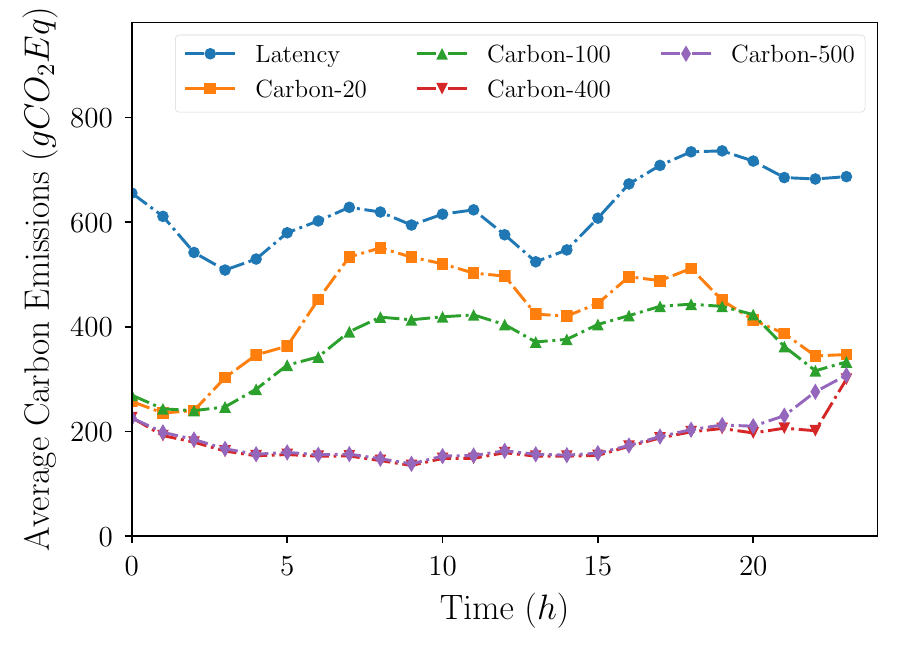} &\\
    (c)  Average Emissions & (d) Hourly Emissions
    \end{tabular}
    \vspace{-0.4cm}
    \caption{\textbf{\emph{Latency and carbon tradeoffs across policies.}}}
    \label{fig:perf_tradeoffs}
    \vspace{-0.5cm}
\end{figure}

Figure \ref{fig:replay_requests}(d) illustrates the resource provisioning across the six AWS regions. The \textit{Latency} policy represents the original provisioning with no redirections. 
As the latency constraints increase, the \systemName initiates re-provisioning of servers from the German ("eu-central-1") region to France due to its lower carbon and lower network latency.
Under \textit{Carbon-100}, a significant portion of the Ohio and Virginia workloads are redirected exclusively to France, as the 100ms latency requirement can be fulfilled. It is worth noting that requests originating from Singapore are only directed to greener locations under the \textit{Carbon-400} policy. This limitation arises from the end-to-end latency from Singapore to any other region surpassing the 100ms threshold.
Moreover, the capacity in France reaches its limit under the \textit{Carbon-400} policy, prompting redirections towards California, in addition to few towards Germany to meet the latency requirements.

\subsubsection*{\textbf{Effects on Carbon and Latency}}

%Figures \ref{fig:perf_tradeoffs}(a)-(d) provides a comparative analysis of all policies. 
%Figures \ref{fig:perf_tradeoffs}(a) and (c) clearly show \systemName's main tradeoff: as the latency constrain relaxes, the potential reduction in emissions increases.
%Although with an average response time of as low as 6ms, the Latency policy yields the highest carbon emissions as requests remain locally, in high intensity regions such as Germany and Singapore. 
%Notably though, as shown by Carbon-20, small latency relaxations can reduce carbon emissions by 25\%.
%Carbon100 is shown to reduce emissions by 37\%, and Carbon-400 reaches the point of diminishing returns with the lowest 70\% reductions in emissions, similarly to Carbon-500 (unrestricted carbon optimization scenario).
%On the other hand, Figures \ref{fig:perf_tradeoffs}(b) and (d) respectively display the hourly variations of average latency and emissions. Compared to the Latency policy, Carbon-20 presents a very small increase in latency of 6ms (while also reducing emissions), and Carbon-100 through Carbon500 shows 5 to 16 $\times$ increases in latency, but with the largest reductions in emissions.

Figures \ref{fig:perf_tradeoffs}(a)-(d) present a comparative analysis of all policies. Figures \ref{fig:perf_tradeoffs}(a) and (c) clearly illustrate the primary tradeoff of \systemName, wherein \textit{the relaxation of latency constraints leads to an increased potential for emissions reduction}. The Latency policy, despite achieving an average response time as low as 6ms, exhibits the highest carbon emissions due to requests remaining localized in high-intensity regions such as Germany and Singapore. Notably, \textit{Carbon-20} demonstrates that even small relaxations in latency constraints can result in a 25\% carbon reductions. \textit{Carbon-100} achieves a 37\% reduction, while \textit{Carbon-400} reaches a point of diminishing returns with a 70\% reduction, similar to \textit{Carbon-500} which represents an unrestricted carbon optimization scenario.
Moreover, Figures \ref{fig:perf_tradeoffs}(b) and (d) display the hourly variations in average latency and emissions, respectively. In comparison to the Latency policy, \textit{Carbon-20} shows a minimal increase in latency of 6ms while simultaneously reducing emissions. \textit{Carbon-100} through \textit{Carbon-500} exhibit latency increases ranging from 5-16$\times$, although delivering the most substantial reductions.
Finally, it is important to note that results would change with other $\alpha$ values. This is primarily due to the fact that \systemName would redirect requests differently due to the trade-off between carbon emissions and the number of servers needed to satisfy latency SLOs.
Specifically, as $\alpha$ increases, \systemName would redirect more requests to greener regions at the cost of latency because this would reduce the number of servers in browner regions. In contrast, as $\alpha$ decreases, \systemName would prioritize latency, opting to handle requests locally despite the carbon costs of setting additional servers.

%The carbon schedulers have lower carbon intensity compared to the replay. When we look at latency comparisons, replay has the best performance with minimum latency averages. 

%\begin{figure}[H]
%    \centering
%    \includegraphics[width=0.4\textwidth]{images/exponential_carbon.png}
%    \caption{Exponential Workload Generation: Carbon Intensity between simulations}
%    \label{fig:exponential_workload_carbon}
%\end{figure}

%\begin{figure}[H]
%    \centering
%    \includegraphics[width=0.4\textwidth]{images/exponential_latency.png}
%    \caption{Exponential Workload Generation: Latency between simulations}
%    \label{fig:exponential_workload_latency}
%\end{figure}

%\begin{figure}[H]
%    \centering
%    \includegraphics[width=0.4\textwidth]{images/bimodal_carbon_intensity.png}
%    \caption{Bimodal Workload Generation: Carbon Intensity between simulations}
%    \label{fig:bimodal_workload_carbon}
%\end{figure}

%\begin{figure}[H]
%    \centering
%    \includegraphics[width=0.4\textwidth]{images/bimodal_latency.png}
%    \caption{Bimodal Workload Generation: Latency between simulations}
%    \label{fig:bimodal_workload_latency}
%\end{figure}

\vspace{-0.25cm}
\section{Related Work}
%The rise of data-intensive jobs and the need for low-latency computing have caused a drift from centralized job execution approaches to geo-distributed job execution. %In centralized job execution, each job is executed within a datacenter, and when data is required from other data centers, the data is collected, aggregated, and brought to a single location. Such an approach quickly becomes impractical as the size of data volumes increases. An alternative approach is to conduct distributed job execution; running jobs closest to the data and aggregating the results at job completion time.
%
%Carbon-aware computing encompasses several key research, some with similar tradeoffs as presented in here~\cite{sukprasert2023quantifying}. 

Recent efforts have concentrated on harnessing the flexibility in energy demand for diverse workloads to diminish their carbon footprint by leveraging the temporal and spatial flexibility of computing~\cite{sukprasert2023quantifying, ecovisor, chasing, embodied2, embodied3, radovanovic2021carbonaware, networks-carbon, chien-carbon, enabling-socc21, lechowicz2023online, wait-awhile}.
Treehouse~\cite{anderson2022treehouse} proposes a software-centric approach to reduce the carbon intensity of datacenter computing by making energy and carbon visible at the application layer. CADRE focuses on carbon-aware data replication to reduce overall carbon footprint, while leveraging load flexibility and interactions with the electricity market to minimize carbon emissions~\cite{xu2015cadre}. 
%Carbon Explorer introduces a framework that considers operational and embodied carbon footprints for renewable energy operation in datacenters. 
\cite{wiesner2021let} investigates the potential of shifting computational workloads to less carbon-intensive periods based on the fluctuating carbon intensity of energy supply.
%
%There has been abundant research in geo-distributed job execution for batch jobs. In such cases, scheduling efficiency is measured using the makespan; the total time required for completion by all the subtasks of a given job, and the average job completion time overall jobs across data centers. Hun et al. \cite{SchedulingJobsAcrossGeoDistributedDatacenters}  improve on the average job completion time of the traditional Shortest Remaining Processing Time (SRPT) scheduling algorithm. It proposes a lightweight “Reordering” add-on and  a new Workload-Aware Greedy Scheduling (SWAG) algorithm, that greedily serve jobs finishing faster by studying the local queues at data centers, achieving as high as 27\% and 50\% improvements in average job completion times over SRPT respectively.
%
%\cite{ALowCarbonKubernetesScheduler} introduces a low-carbon extension to the Kubernetes default scheduler, that sorts serving cloud regions by carbon intensity and migrates workloads to regions with lower carbon cost. However, the proposed framework is evaluated on BOINC (Berkeley Open Infrastructure for Network Computing) jobs; which are primarily batch jobs. On the other hand, numerous works like Trevino-Martinez et al.\cite{EnergyCarbonFootprintOptimizationInSequenceDependentProductionScheduling} have employed integer programming techniques to devise new frameworks for low-carbon scheduling.
%
\cite{ALowCarbonKubernetesScheduler} introduces a low-carbon extension to the Kubernetes scheduler, sorting cloud regions by carbon intensity and migrating workloads to regions with low carbon cost. However, the proposed framework is evaluated primarily for batch jobs. On the other hand, numerous works have employed integer programming techniques to devise new techniques for low-carbon scheduling~\cite{EnergyCarbonFootprintOptimizationInSequenceDependentProductionScheduling}.
Carbon-aware geo-distributed scheduling is particularly relevant for Machine Learning (ML) workloads requiring long periods of execution \cite{CO2EmissionAwareSchedulingForDNNTrainingWorkloads}. %Therefore, the development of greener scheduling is an important research direction towards making ML training more sustainable. 
%Haghshenas et al.\cite{CO2EmissionAwareSchedulingForDNNTrainingWorkloads} exploits the iterative nature of ML training, and designs a framework to both shift workloads in time and between locations between epochs, reducing emissions by 13-57\%. 
%Zhang et al. \cite{zhang2023sustainable} propose a multi-agent reinforcement learning method to learn an optimal scheduling strategy by interacting with a cloud simulation with real ML workloads, improving GPU utilization by 28.6\% while using less energy and carbon. They \cite{zhang2023sustainable} take a different approach, and focus on delay-tolerant workloads. 
\cite{zhang2023sustainable} proposes Cucumber, an admission control policy that leverages load and energy forecasting techniques to determine scheduling strategies to use renewables. 
Unlike the previous works, \systemName is the first  framework that seamlessly integrates server provisioning and request scheduling for a geo-distributed web application, with a particular focus on interactive web requests. 

\vspace{-0.3cm}
\section{Conclusion and Future Work}
%Through empirical evaluation of \systemName through the CAP and CAS models, we observed significant reductions in the carbon footprint of web traffic in data centers by up to 74\%. This research represents an important step in  carbon-saving efforts for geo-distributed web applications. However, there is still scope for performing more thorough analysis and exploring additional strategies for making the system more efficient.
%In particular, as future work an auto-scaler that monitors resource usage in all regions and dynamically adjusts the resource allocation. Empirical analysis can be performed on auto scaling algorithms to enhance the overall carbon-saving capabilities of the system. By continuing to refine upon these algorithms, we can take a significant step towards building a greener future for the cloud community.

This paper introduced \systemName, a carbon-aware scheduler and provisioner designed for distributed web applications. 
%We implement \systemName as a Kubernetes scheduler, and our results highlight its significant potential in achieving considerable reductions in carbon footprint for web applications.
%By using several latency and carbon-aware policies, we show that the adaptation of performance constraints leads to substantial potential for carbon reductions. 
At the heart of \systemName lies a multi-objective optimization that minimizes both resources and latency, introducing a novel method to control the load balancing of web applications.
We observe substantial savings in the carbon footprint, reaching up to 70\% with controllable and negligible losses in performance while meeting all SLOs. 
\systemName represents a crucial advancement in carbon-aware schedulers for distributed and geo-distributed applications. Further analysis and exploration of additional spatiotemporal carbon-aware strategies are warranted to enhance the system's efficiency.
As a potential avenue for future work, the implementation of auto-scaling policies that continuously monitors resource utilization across regions to dynamically adapt allocations could be explored. %The analysis of such auto-scaling algorithms can further improve the overall carbon-saving capabilities presented with \systemName. 

\begin{acks}
We thank the reviewers for their valuable comments, which improved the quality of this paper, and Electricity Maps for access to their carbon-intensity data. This research is supported by NSF grants 2211302, 2211888, 2213636, 2105494, and Army contract W911NF-17-2-0196.
\end{acks}

\bibliographystyle{ACM-Reference-Format}
\bibliography{main.bib}

\end{document}